# Femtosecond Hot-Exciton Emission in a Ladder-Type π-Conjugated Rigid-Polymer Nanowire

D. C. Dai* and A. P. Monkman

*Department of Physics, Durham University, South Road Durham DH1 3LE, United Kingdom*
e-mail: dechang.dai@dur.ac.uk

(The first version finished on Oct. 2011, present version finalized on March 2012.)

A hot-exciton is usually the initial elementary excitation product of the solid phase, particularly in low dimensional photonic materials, which is a bottle-neck to all subsequent processes. Measurement of hot-exciton emission (HExEm) is a great challenge due to fast $E_K$ relaxation and thus very weak transient emission. Here we report the first unambiguous observation of femtosecond HExEm from thin films of a model *quasi*-one-dimensional π-conjugated organic rigid-rod quantum nanowire, MeLPPP (methyl-substituted ladder-type poly(*para*-phenylenes), by using femtosecond time-resolved fluorescence spectroscopy. The results show the clear HExEm from the cooling hot-excitons has a lifetime of ~500 to ~800 fs, and concomitant very weak density-dependent singlet-singlet annihilation (SSA) due to this ultrashort dwelling time.

Excitons are the elementary excitation in the solid phase, and the primary electronic excited state in low dimensional photonic materials for (organic) light emitting device (LED), micro-cavity laser and photovoltaics (PV) *etc.*, such as polymer, nanotube, $C_{60}$ and graphene, quantum well, wire and dot [1-6]. Compared to the molecular (Frenkel) and crystalline (Wannier-Mott) excitons, (*quasi*-) one-dimensional (1D) excitons are of intermediate form, which is always bound to a wire, but may form across two wires. The strong quantum confinement effect and electron-phonon coupling stabilize these excitons so that they are dominant from low to high temperatures [1-6].

A hot-exciton [7] is usually the initial excitation product caused by photogeneration or charge recombination for example, it is considered as an exciton with kinetic energy $E_K$ that considerably exceeds the mean thermal energy $k_B T_L$, i.e., $E_K > k_B T_L$, here $k_B$ is the Boltzmann constant, and $T_L$ refers to the lattice temperature of the solid.

The current picture of hot-exciton evolution immediately after creation is very fast $E_K$ relaxation via internal-conversion (cooling) [6,8], and subsequent processes such as excitation energy transfer, migration and trapping, annihilation, dissociation into polaron pairs or charge-separated states, inter-system-crossing of singlet to triplet, formation of bi-exciton, electron-hole plasma (EHP), and Bose-Einstein condensate (BEC). Finally, luminescence is given from recombination of the relaxed (cooled) excitons on the lowest energy states at the bottom of density of states (DOS), as illustrated in Fig. 1. All of these are determined by the basic properties of materials used [1-6,8]. Thus the hot-exciton is a bottle-neck for the divergent consequences of exciton evolution, therefore investigation of hot-excitons has great fundamental importance.

Hot-exciton emission (HExEm), arising from the radiative recombination of hot-excitons, is a transient phenomenon, it gives the most direct information on $E_K$ distribution and evolution [7,8]. Since a hot-exciton is thermal non-equilibrium state of the system, the measurement of HExEm is a great experimental challenge due to very fast cooling, i.e., a hot-exciton always tends to cool rapidly by emission of optical and acoustic phonons towards an (*quasi*-) equilibrium state with lower energy. Little direct observation has been convincingly made because of this, and all of previous claims have been done in the frequency domain or by confusion of resonant Raman scattering (RRS) [7,8]. The very recent reports of tailored HExEm [9,10] are still in the frequency domain, which utilizes a very strong quantum confinement effect in combination with a cavity effect. However, an identification of HExEm in the time domain has never been seen. Here we report the first unambiguous femtosecond HExEm from a model π-conjugated rigid-rod organic *quasi*-1D quantum nanowire, MeLPPP (methyl-substituted ladder-type poly(*para*-phenylenes), using femtosecond time-resolved fluorescence up-conversion spectroscopy.

The evolution of excitons in polymers has attracted extensive studies using time-resolved fluorescence spectroscopy in combination with steady state spectroscopy, to date two regimes have been investigated: (i) the dephasing and migration of coherent excitons on femtosecond timescale, for example, a dephasing time $T_2^* = $ ~250 fs has been recently described in non-rigid poly(*para*-phenylene vinylene) (PPV) [11,12]; (ii) all other studies are mainly focused on the spectral shift of luminescence $\lambda_{em}$ from the migrating relaxed excitons on the picosecond to nanosecond timescale [13-21]. Obviously, there is a clear gap between (i) and (ii): in the frequency domain this gap appears as a large shift from the excitation $\lambda_{ex}$ to the onset of $\lambda_{em}$, as shown in Fig. 1, which corresponds to initial hot-exciton cooling [6]. If excitation occurs yielding a large $E_K$, any resultant emission from the hot-excitons will lie underneath the strong linear absorption band, therefore its intensity will be very weak and it is usually not expected to be observed in experiment [1-8,11-21]. In the time domain this gap represents a period between $T_2^*$ and ~3 ps. To date no one has explored the physical processes of hot-exciton in π-conjugated polymers because of this [1-6, 11-21].

Femtosecond pump-probe techniques are of little use here due to the superposition of stimulated emission (SE) from the hot-exciton with photoinduced absorption (PA) to high-lying excited states [2-5,14,22-25]; there are limitations on the availability of the deep-blue spectral component from the white-light supercontinuum as a probe [22-25].

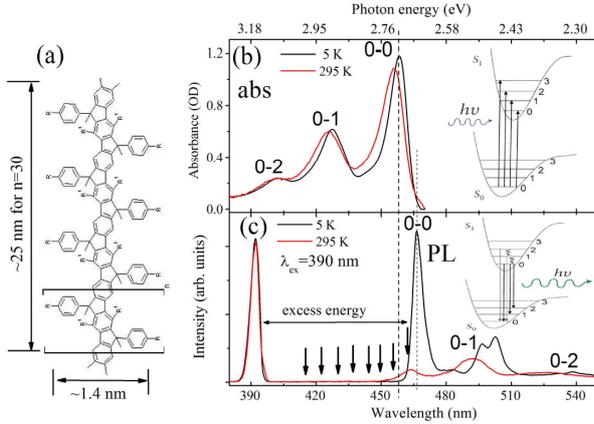

FIG. 1 (color online). Chemical structure of a rigid-rod MeLPPP nanowire (a), absorption (abs) (b) and photoluminescence (PL) (c) spectra of thin film. R-$C_{10}H_{21}$, and $R^1$-$C_6H_{13}$. The downward arrows in (c) indicate the emission wavelengths ($\lambda$) monitored. The insets in (b) & (c) show the illustrating potential diagrams of energy levels and the corresponding optical transitions.

In pristine thin films, π-conjugated polymers are typically amorphous glasses, having a statistical distribution of chain lengths and random orientations, and each chain usually is composed of multiple segments due to chain folding, twisting, kinks or chain defects *etc*., all of these yield poorly resolved absorption and emission spectra and provide a large heat bath to enable excitons to migrate non-dispersively over relative large distance during their lifetimes [1-6,11-25]. Moreover, the rich vibronic modes and the strong electron-phonon coupling on each single chain greatly broaden the overall optical transition cross-sections in absorption and emission [1-6, 11-25]. These all conspire to produce large degrees of homogenous and inhomogeneous broadening with concomitant very short $T_2$ and $T_2^*$ respectively, usually on the scale of a few hundred femtosecond or less [11,12]. Whereas ladder-type PPP is a class of rigid-rod *quasi*-1D organic quantum wire with discrete chain sizes [26], as a result of inter repeat unit methylene bridges, see Fig. 1 for structure. They can be easily synthesized at high purity (>99%) and low polydispersity of chain length. All of these features have lead to the observation of nearly prefect symmetric spectral line shapes as in molecular spectroscopy [26,27], which is shown in Fig. 1, and the estimated longest $T_2^*$ =~520 fs for a single chain at low temperature [28]. Therefore MeLPPP is an ideal model representing polymer chains and rigid inorganic nanowires, for investigation of exciton dynamics.

The pristine films of MeLPPP (purity >99.5%, $M_n$=25 kDa, *approx.* n=30, polydispersity 1.3) were made by spin-coating onto sapphire substrate from its toluene solution, 15 mg/ml. The typical thickness is ~125 nm. Samples are mounted in a Helium closed-cycle cryostat for both 5.0 K and room temperature measurements. The chemical structure, optical absorption and photoluminescence (PL) spectra are shown in Fig. 1. The PL quantum yields estimated at very low excitation power ($P$) are ~50% and ~25% at 5 K and room temperature, respectively.

The experimental setup of fs time-resolved fluorescence up-conversion spectroscopy has been previously described in ref. [29], which is similar to others in reports [13-17,20]. The typical cross-correlation time of the system is $\Delta t$=360 fs, which is shown in Fig. 2(a) as the $t_0$ pulse. A bandpass filter (FF01_447/60, Semrock) is used to fully block the excitation scattering at $\lambda_{ex}$=390 nm (3.18 eV).

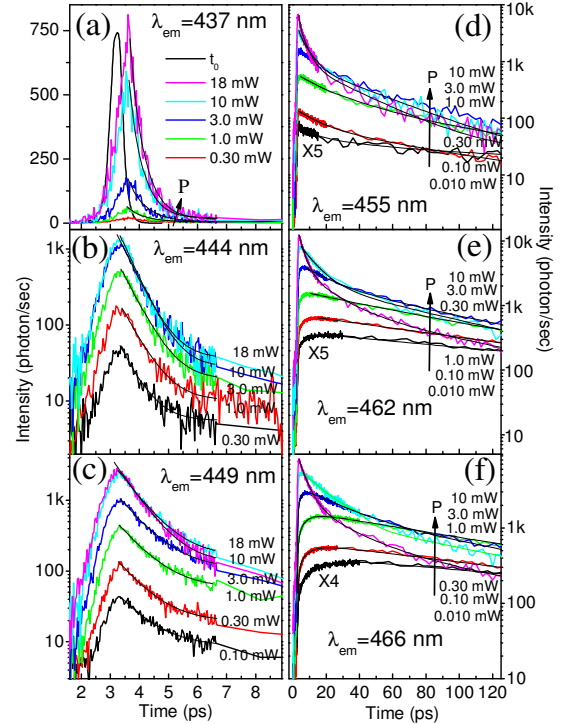

FIG. 2. (color online) Excitation power ($P$) dependent ultrafast emission dynamics of the hot and cooled excitons in MeLPPP at 5 K. The thin line on each curve is the exponential fitting, and the arrows with $P$ indicate the increase of excitation power, these apply to the subsequent FIG. 3.

The spectra in Fig. 1, with very small Stokes shift indicative of a true rigid-rod chain [26], represent Franck-Condon transitions for excitonic states in MeLPPP, the excitation laser with a $\tau_{ex}$=~280 fs directly creates a hot-exciton population $N_\lambda$ in the first electronic excited state manifold $S_1v_{2,j}$, (j signifies a low energy chain breathing mode [27]) with an excess energy of ~0.47 eV with respect to the 0-0 absorption maximum ($S_0v_0 \rightarrow S_1v_0$) at ~457 nm (2.71 eV). This is at least twice the energy of C=C vibronic modes (~1600 $cm^{-1}$ =~0.20 eV) [27]. Here $N_\lambda$ is proportional to $P$.

Figure 2 depicts the wavelength ($\lambda$) and $P$ dependent ultrafast dynamics of exciton emission in MeLPPP film at 5 K. The emission at $\lambda$=437 nm (2.84 eV), 444 nm (2.79 eV) and 449 nm (2.76 eV) in Fig. 2(a)-(c) all exhibit weak but clear peaks, which can not be observed from a bare sapphire substrate only. These lie deep beneath the absorption band and correspond to the radiative transition $S_1v_{1,j} \rightarrow S_0v_0$, and are therefore interpreted to be HExEm from the cooling hot-excitons. Their emission profiles are all similarly pulse-like and have a clear average time delay $\tau_d = \sim 0.32$ ps with respect to $t_0$ (shown in Fig. 2(a)), this $\tau_d$ marks a period of initial $T_2^*$ process and internal conversion from $S_1v_{2,j}$ to $S_1v_{1,j}$. These pulse widths (full width at half maximum, FWHM) are all typical $\tau_W = 1.0 \pm 0.10$ ps at $P = 0.30$ mW, slightly shorten to $0.80 \pm 0.10$ ps at 10 mW for both 437 nm and 444 nm, and $1.0 \pm 0.10$ ps at 449 nm. These values are appreciably longer than the width of $t_0$, $\Delta t = 360$ fs, indicating that the time resolution of our system is sufficiently short to resolve these ultrashort emission features. The pulse-like peaks of HExEm shown in Figs. 2 and 3 are on a logarithmic intensity scale except Fig. 2(a), are cross-correlation traces between the real HExEm (duration $\tau_{HExEm}$) and the up-conversion gating pulse ($\tau_p = \sim 180$ fs), therefore $\tau_{HExEm}$ can be calculated by ($\tau_W$-$\tau_p$), thus $\tau_{HExEm} = \sim 500$ fs to $\sim 800$ fs, is appreciably longer than the $\tau_{ex} = \sim 280$ fs. As the observation window moves from 437 nm to 449 nm the decays become more developed with well resolved trailing edges and all show excellent fit to a single exponential function: the lifetimes at $\lambda = 437$ nm and 444 nm are from $\tau = 0.60 \pm 0.05$ ps at $P = 0.30$ mW to $0.50 \pm 0.05$ ps at 10 mW, whereas at $\lambda = 449$ nm $\tau$ is slightly longer, from $0.85 \pm 0.05$ ps at 0.30 mW to $0.75 \pm 0.05$ ps at 10 mW, this is attributed to the contribution of HExEm from the exitons on the upper $S_1v_{0,j}$ states. At each $\lambda$ the values of $\tau_W$ and $\tau$ at high $P$ are slightly short than low $P$, this is attributed to the very weak excitonic singlet-singlet annihilation (SSA) effect, which slightly decrease the exciton density $N_\lambda$ thus $\tau$ as $P$ increase.

At $\lambda = 455$ nm (2.73eV) we are nearly at the 0-0 absorption maximum at $\sim 457$ nm and the onset of $\lambda_{em}$ (Fig. 1), $E_K$ is rather small and thus HExEm is not expected, but also the emission dynamics in Fig. 2(d) show clearly different behaviours from the previous $\lambda$: the curve at low $P$ ($<0.30$ mW) fits well to a bi-exponential decay, a fast $\tau_2 = 10 \pm 1$ ps and a slow $\tau_1 = 75 \pm 5$ ps, these emergent very long lifetimes are indicative of emission from the fully cooled excitons $N_\lambda$ migrating towards the bottom of DOS. At moderate $P$ (0.30 mW$<P<$3.0 mW), there are the clear first signs of fluorescence lifetime quenching due to the onset of SSA on increased $N_\lambda$. At $P>3.0$ mW a very fast lifetime $\tau_3 = 0.70 \pm 0.05$ ps appears, indicating the growing efficiency of SSA which effectively outcompetes all other decay channels at very high $N_\lambda$. This should not be confused with the previous HExEm from the cooling hot-excitons. These results show the excitons $N_\lambda$ become 'trapped' on these relatively high energy chains and the SSA becomes effective with the enormously increased dwell time $\tau$ of the excitons on these chains.

At $\lambda = 462$ nm (2.68 eV), the blue edge of $S_1v_{0,j}\rightarrow S_0v_0$ emission band (Fig. 1), the possibility of HExEm can be completely ruled out, and a clear build-in is observed immediately after $t_0$, see Fig. 2(e), indicating the cooled excitons $N_\lambda$ migrating into this detection window. A single exponential growth fit gives a lifetime from $\tau_{gr} = 9.0 \pm 1$ ps at $P = 0.010$ mW to $5.0 \pm 0.5$ ps at 0.30 mW, in agreement with the fast decay $\tau_2$ at 455 nm, and indicative of increasing migration induced filling of lower energy states. Above 1.0 mW SSA tends to dominate the decays therefore build-in disappears. At low $N_\lambda$, the decay is also of typical single exponential, giving a $\tau_1 = 160 \pm 10$ ps at 0.010 mW. Some quenching to $120 \pm 10$ ps at 0.10 mW and 0.30 mW due to the weak SSA is seen. As $P>0.30$ mW, a $\tau_2 = 31 \pm 10$ ps can be resolved and shortens to $7.0 \pm 1$ ps as $P>1.0$ mW, also consistent with $\tau_2$ at 455 nm. Again at high $P$ ($>10$ mW), an ultrashort $\tau_3 = 1.0 \pm 0.1$ ps emerges, indicative of efficient SSA not HExEm.

At $\lambda = 466$ nm (2.66 eV), the onset of absorption and the peak of $S_1v_{0,j}\rightarrow S_0v_0$ transition, the emission emanates primarily from the fully cooled and immobilized excitons $N_\lambda$ that have migrated to the bottom of DOS. Concomitant with this a long population accumulation (build-in) process spans $>30$ ps at $P = 0.010$ mW, having an exponential growth of $\tau_{gr} = 12 \pm 2$ ps, which gradually shortens to 3.0 ps at 1.0 mW (Fig. 2(f)) due to the increasing filling rate at high $N_\lambda$. As $P>3.0$ mW the build-in disappears due to SSA again. The single exponential decays at 0.010 mW and 0.10 mW give $\tau_1 = 200 \pm 10$ ps and $180 \pm 10$ ps, respectively, representing the longest fluorescence lifetime measurable. At 0.30 mW the curve turns bi-exponential, indicative of very slow migration further down the DOS at high $N_\lambda$ again. At 10 mW, again SSA starts to dominate giving $\tau_3 = 1.0 \pm 0.1$ ps, this is all consistent to 462 nm and 455 nm.

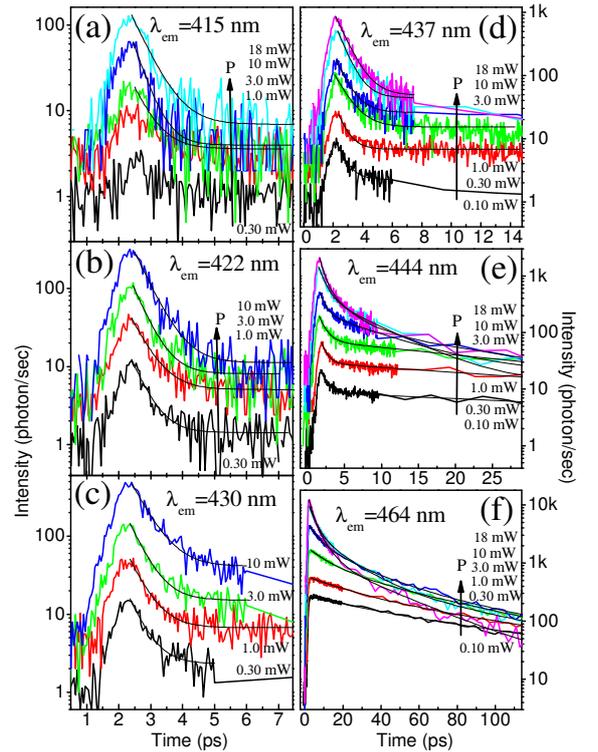

FIG. 3. (color online) $P$ dependent ultrafast emission dynamics of the hot and cooled excitons in MeLPPP at room temperature.

In contrast to 5 K, given $k_BT_L = \sim 26$ meV at room temperature, there is a large amount of thermally activated vibronic and chain breathing modes in MeLPPP, and hence a faster $T_2^*$ and broadened $E_K$ distribution [7, 27]. The spectral peaks and onset of $\lambda_{em}$ all undergo blue-shift, the PL quantum yield is greatly reduced to $\sim 25\%$, particularly the 0-0 emission is much weaker than the 0-1 emission primarily due to self-absorption (Fig. 1). These results are detailed in Fig. 3.

The weaker but still clear HExEm at $\lambda$=415 nm (2.99 eV), 422 nm (2.94 eV), and 430 nm (2.88 eV), all have similar pulse widths, $\tau_W$= ~0.80 ps (0.70-0.90 ps), and same decays, $\tau$=0.45±0.05 ps (Fig. 3(a)-(c)). Within experimental uncertainty these are the same as at 5 K. At 437 nm (2.84 eV), $\tau_W$ lengthens to 1.0±0.10 ps, and the lifetime to $\tau$=0.60 to 0.80±0.05 ps (Fig. 3(d)), indicating a slight slowing down of hot-exciton cooling than previous $\lambda$, again in line with what we see at 5 K pre-fluorescence onset.

The blue-shift of the onset of $\lambda_{em}$ to 444 nm (2.79 eV), is further indicated by bi-exponential decays in Fig. 3(e): a slow $\tau_1$=50±10 ps indicates some cooled excitons dwelling on these sites, a fast $\tau_3$=1.0±0.10 ps sitting on the slow component, which is independent of $P$ and in consistent to $\lambda$=437 nm in Fig. 3(d), is clearly HExEm.

At longer $\lambda$, the HExEm disappears and the migration component is seen again. At 464 nm (2.67 eV) in Fig. 3(f), the 0-0 emission peak, a slow $\tau_1$=85±10 ps at 0.10 mW and 0.30 mW, shortens to 60 ps at 10 mW, this is much shorter than $\tau_1$=200 ps at 5 K and is ascribed to thermally enhanced quenching, which also causes the large reduction on PL quantum yield. A fast $\tau_2$=20±5 ps at 0.10 mW and 15 ps at 10 mW, indicates the excitons are still actively migrating to the bottom of DOS, but the slow build-ins are not observed, indicative of being at thermal equilibrium and thus non-dispersive migration. A $\tau_3$=9.0±1.0 ps emerges at 1.0 mW and shortens to ~3.0 ps as $P$>10 mW, resulting from SSA effect which is enhanced by thermally activated hopping.

The initially created coherent hot-excitons $N_\lambda$ become incoherent within $T_2^*$=~520 fs [28] most probably by emission of lower energy phonons than the chain breathing mode at 113 cm$^{-1}$ (14 meV) [27], which has a vibrational period of 295 fs very possibly corresponding to the observed $\tau_d$=~0.32 ps.

The HExEm emanating at $\lambda$<437 nm can not be from any residual oligomers or segments, given that (i) the emission peak of an $n$=11 chain is at 449 nm [30], and a pentamer ($n$=5) is at 437 nm [31], (ii) the femtosecond $\tau$ values can not be explained by the Forster energy transfer induced quenching effect: the estimated efficiency needed to get such rapid transfer, $E$=~99.6% is far too high, and the estimated distance $r$=~2.0 nm doesn't match the minimum chain separation due to its size shown in Fig. 1a and the side-chain spacing effect (Supplemental Material).

HExEm is easy to confuse with RRS at time $t_0$ [7], here RRS can be completely ruled out as it can not explain any of the observed features: the $\tau_d$= ~0.32 ps, the broader $\tau_W$= ~0.80 ps than the $\tau_{ex}$=~280 fs and $\Delta t$=360 fs, the gradually increasing emission intensity and lengthening decays with increasing $\lambda$ at a fixed $P$, the small nonlinear amplitude increase with increasing $P$, and the temperature effect. Thus, we conclude the unambiguous observation of femtosecond HExEm from MeLPPP nanowire. It is not possible to resolve which vibronic modes contribute to the HExEm as the vibronic manifold is complex with many overtones from low energy modes [27]. Whereas the weak HExEm suggests a tiny fraction of $N_\lambda$ on the $S_1v_{1,j}$ states recombine directly whilst the bulk relaxes to the $S_1v_{0,j}$ states during the first ~1.0 ps immediately after $t_0$.

From our data, the exciton migration clearly takes place immediately after the hot-exciton cooling, and depends on $N_\lambda$, $\lambda$ and $k_BT_L$. The increasing decays ($\tau_1$ & $\tau_2$) with $\lambda$ is a clear indication of dispersive singlet exciton migration [13-25], confirmed by the increasing $\tau_{gr}$ with $\lambda$ at 5 K. $k_BT_L$= ~26 meV at room temperature straightforwardly means a significant Boltzmann population on the low energy phonon modes, hence the migration starts at shorter $\lambda$ and we do not observe the build-in of $\lambda_{em}$ as at 5 K.

And, our results also indicate that the SSA is determined by exciton density $N_\lambda$, which is equivalent to an average exciton-exciton separation, and is more sensitive to $k_BT_L$, and the dwelling time $\tau$ on the sites at $\lambda$, this allows excitons to find quenching sites throughout the film, thus explaining the much lower PL quantum yield at room temperature than at 5 K and the very weak SSA in HExEm.

We thank Prof. Ullrich Scherf for supplying high-purity MeLPPP polymer and fruitful discussions.

Figure captions

FIG. 1 (color online). Chemical structure of a rigid-rod MeLPPP nanowire (a), absorption (abs) (b) and photoluminescence (PL) (c) spectra of thin film. R- $C_{10}H_{21}$, and $R^1$- $C_6H_{13}$. The downward arrows in (c) indicate the emission wavelengths ($\lambda$) monitored. The insets in (b) & (c) show the illustrating potential diagrams of energy levels and the corresponding optical transitions.

FIG. 2. (color online) Excitation power ($P$) dependent ultrafast emission dynamics of the hot and cooled excitons in MeLPPP at 5 K. The thin line on each curve is the exponential fitting, and the arrows with $P$ indicate the increase of excitation power, these apply to the subsequent FIG. 3.

FIG. 3. (color online) $P$ dependent ultrafast emission dynamics of the hot and cooled excitons in MeLPPP at room temperature.

Fig. 1

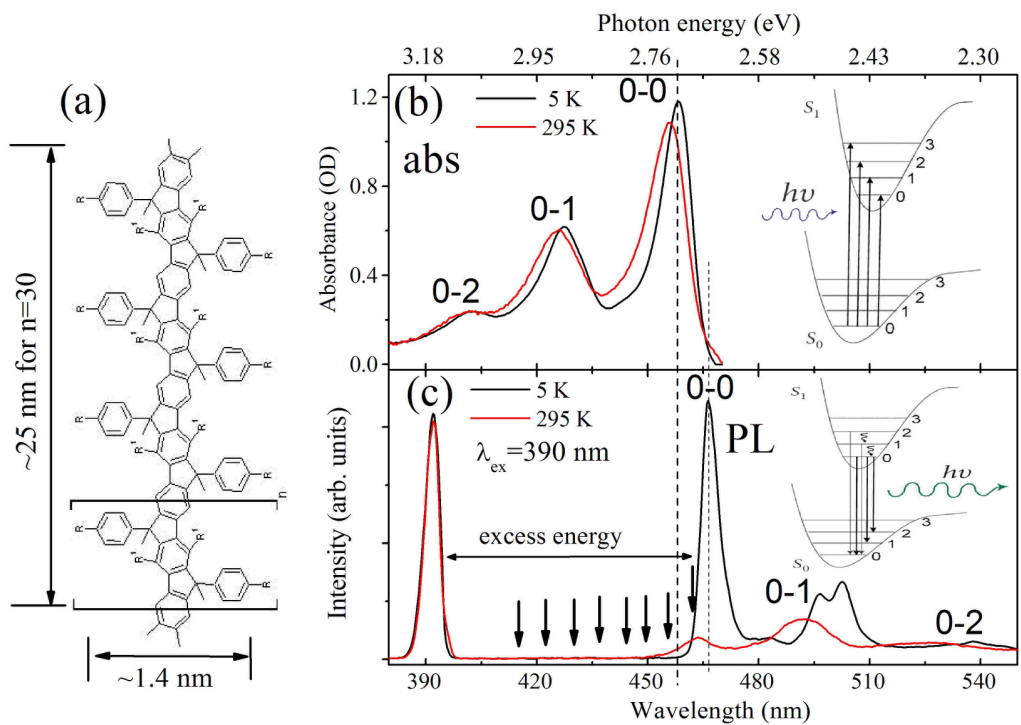

Fig. 2

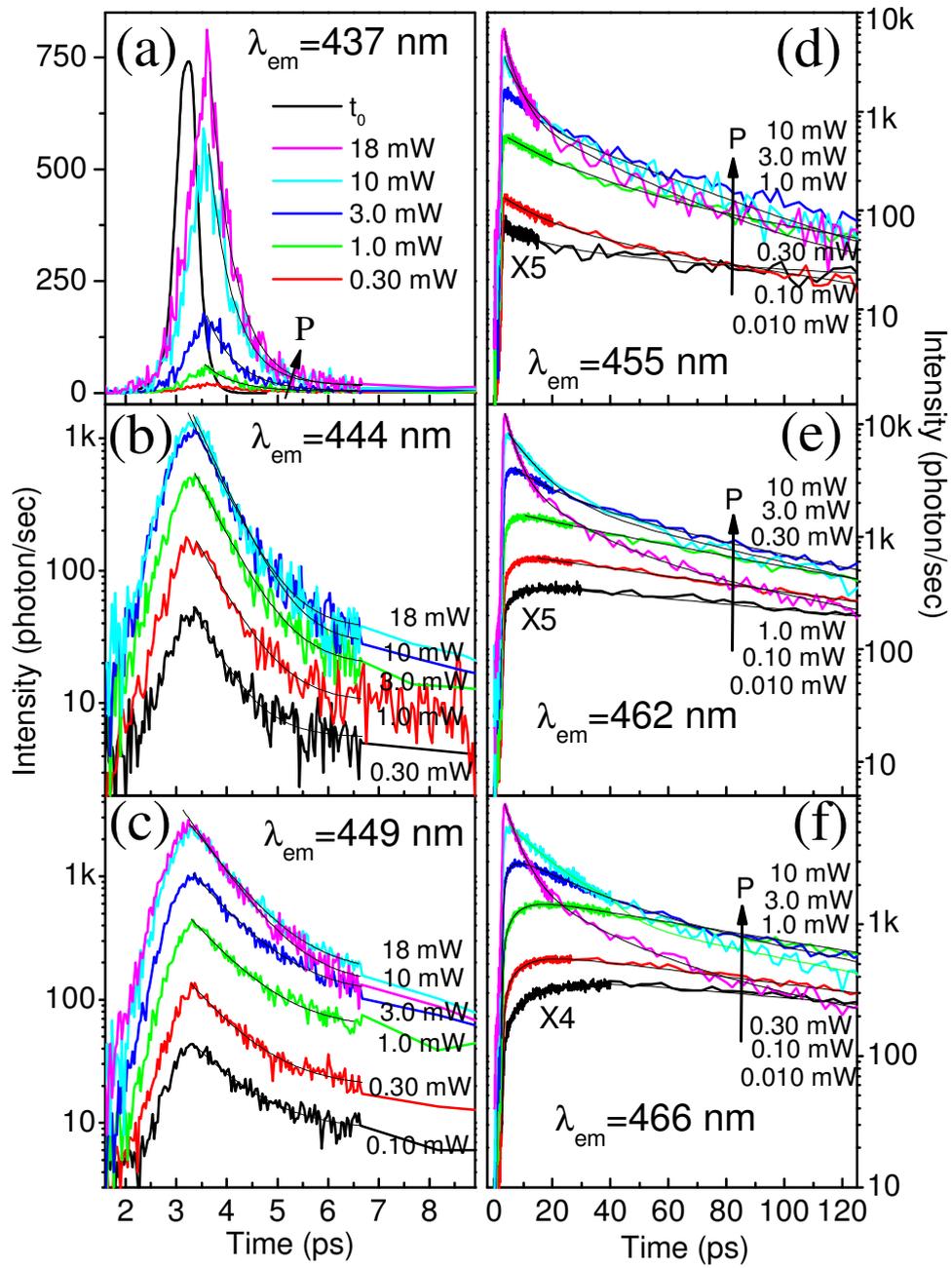

Fig. 3

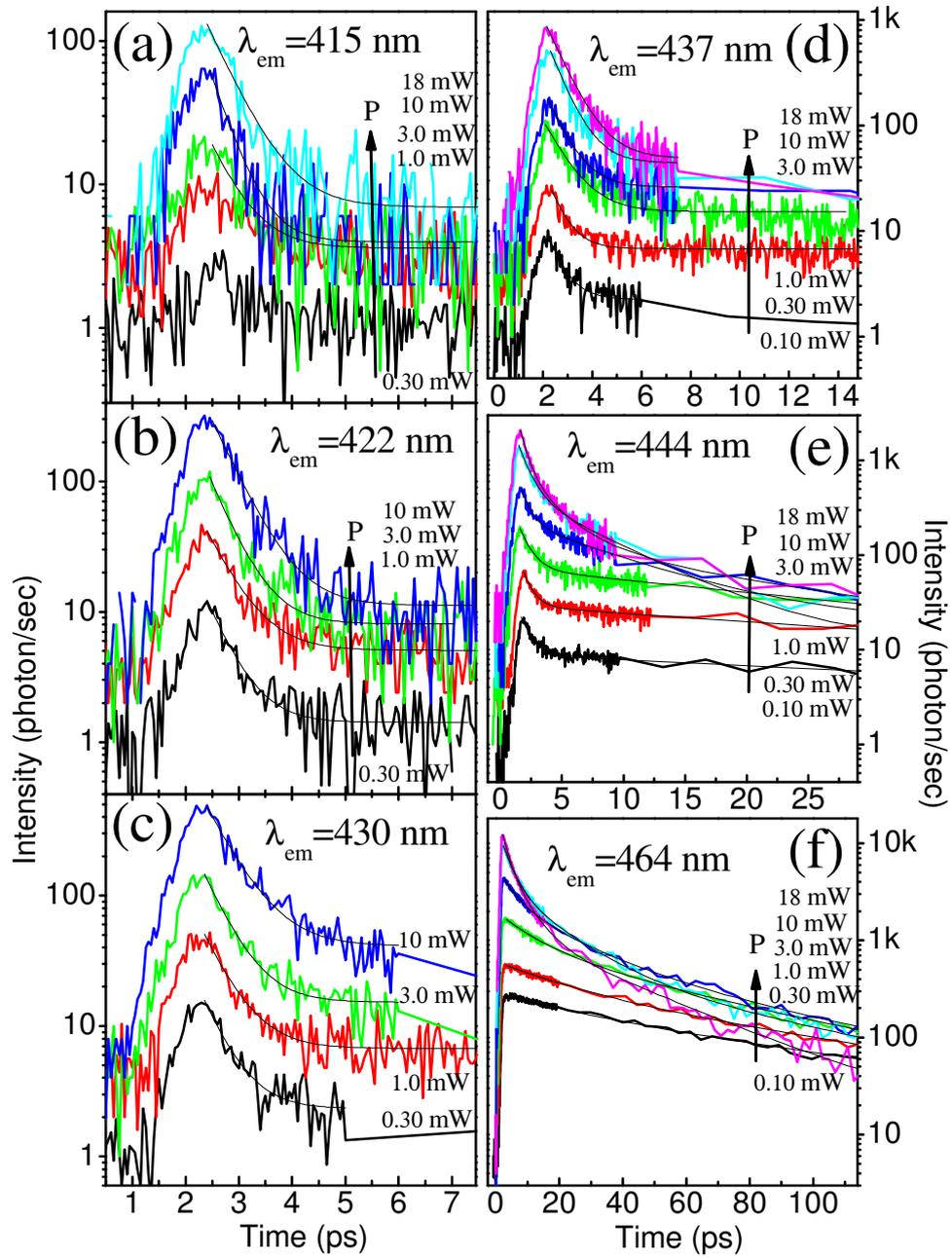

Supplemental Material for

# Femtosecond Hot-Exciton Emission in a Ladder-Type π-Conjugated Rigid-Polymer Nanowire

D. C. Dai* and A. P. Monkman

*Department of Physics, Durham University, South Road Durham DH1 3LE, United Kingdom*
e-mail: dechang.dai@dur.ac.uk

**This file includes:**

**1. Test of excitation laser scattering and blocking by optical filters.**

Figs. S1, S2.

**2. Supplementary data to Fig. 3 in the main text.**

Fig. S3.

**3. Summary of fitting results by exponential functions to the data in Figs. 2 & 3.**

**4. Confusion over femtosecond HExEm decay and analysis by Forster-Dexter energy transfer model.**

Fig. S4.

**References and Notes**

**1. Test of excitation laser scattering and blocking by optical filters** in femtosecond time-resolved fluorescence up-conversion spectroscopy.

The excitation scattering ($\lambda_{ex}$=390 nm) has been found to bring much distortions to the experimental results. Here we use a specific filter, FF01_447/60 (Semrock) to fully block the scattering, some test data are shown below in Figs. S1 & S2.

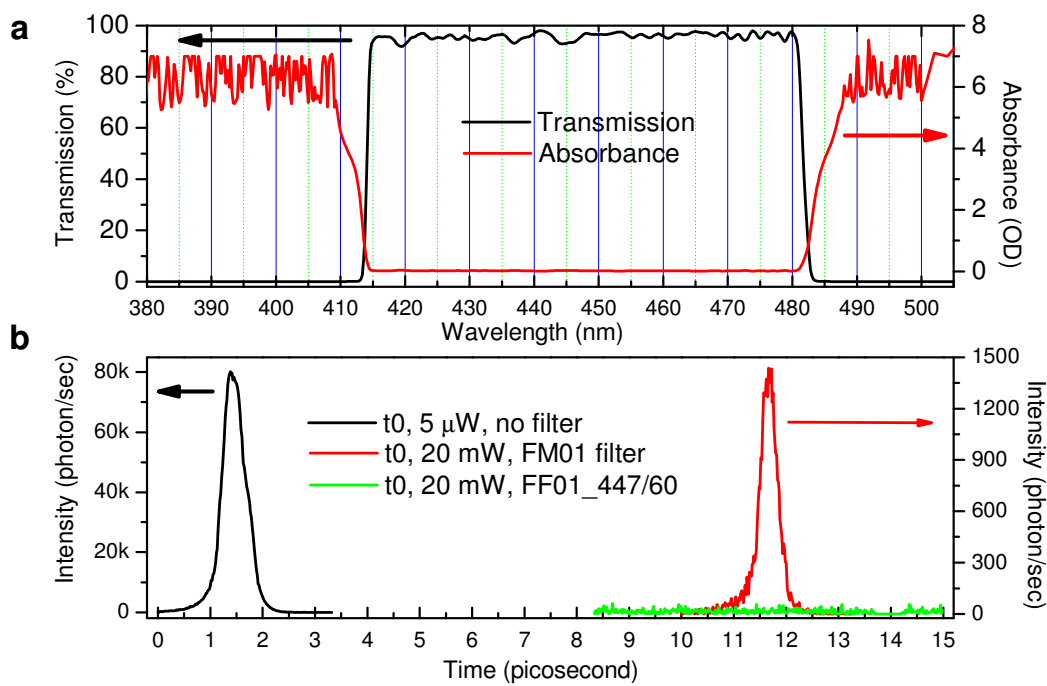

**FIG.** S1. Blocking efficiency of excitation scattering by the filter FF01_447/60. **a,** Transmission and absorbance spectra of the filter, showing an effective transmission window between 413 nm and 482 nm. **b,** Comparison of $t_0$ pulses recorded as the system response function, showing the excitation scattering (max. 80k at 5 μW ) can be efficiently blocked by the filter FM01 from Thorlabs (red curve, max. 1400 at 20 mW), and completely by FF01_447/60 (green curve, intensity 0 at 20 mW). The time delay of red pulse is induced by the thickness of filter, both filters have similar thickness in mm.

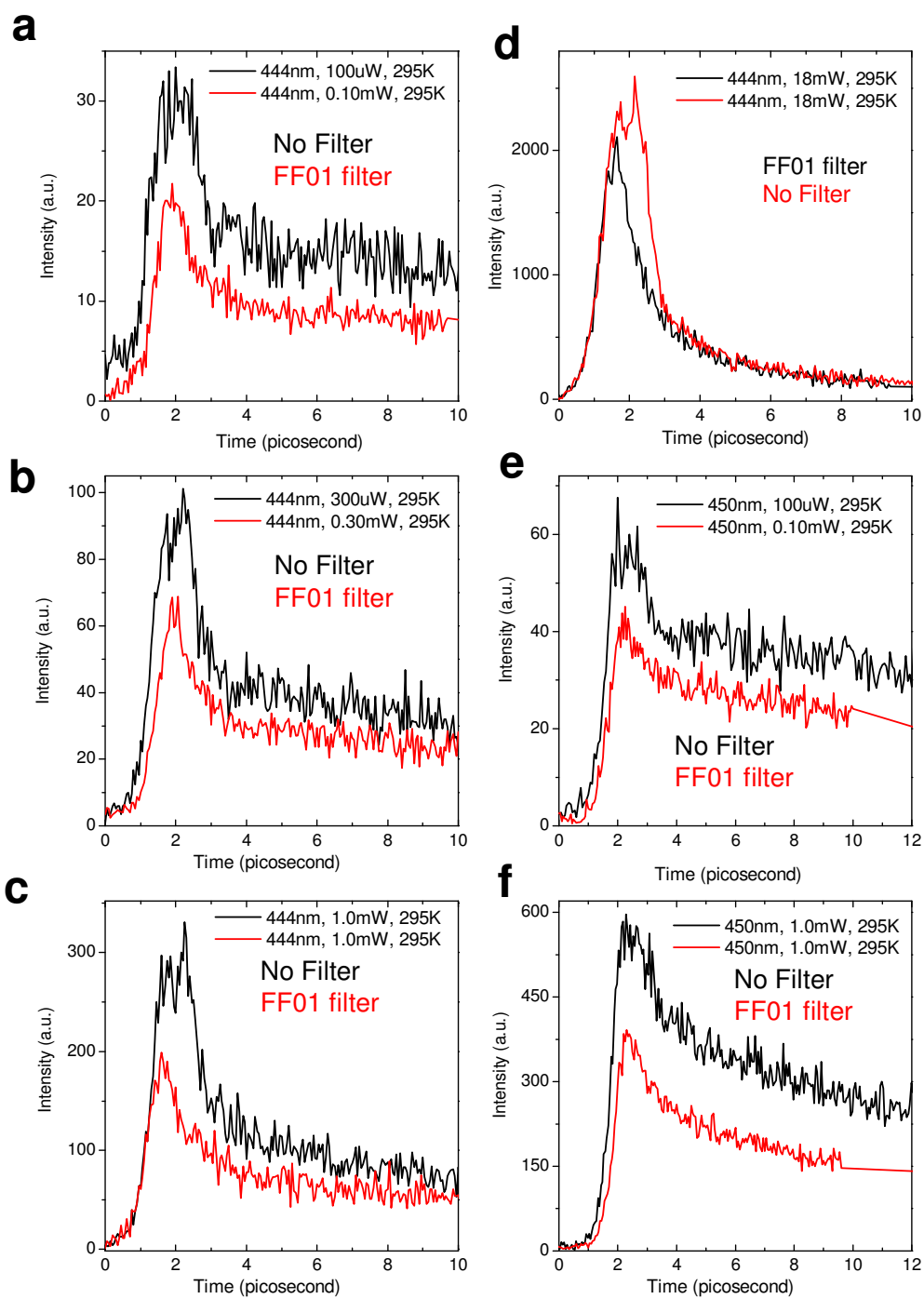

FIG. S2. Comparison of experimental data with and without filter FF01. **a-f**, showing apparent distortions within the first few picoseconds induced by excitation scattering.

## 2. Supplemental data to FIG. 3 in the main text.

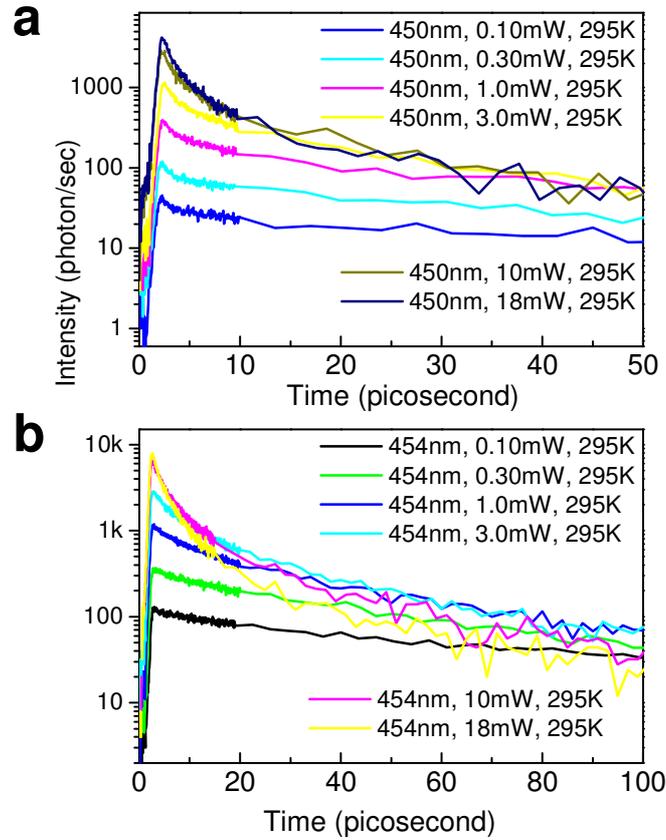

FIG. S3. *P* dependent ultrafast emission dynamics from migrating relaxed excitons in MeLPPP film at room temperature. **a**, At this $\lambda$ the HExEm as $\lambda_{em}$=444 nm in Fig. 3(e) disappears, the ultrafast relaxation, exciton migration and *P* dependent SSA are clearly shown. **b,** The lengthened decays are similar to $\lambda_{em}$=464 nm in Fig. 3(f).

## 3. Summary of fitting results by exponential functions to the data in Figs. 2 & 3.

| **437 nm, 5K** | 0.30 mW | 1.0 mW | 3.0 mW | 10 mW | 18 mW |
|---|---|---|---|---|---|
| Exp decay | 0.61 ps | 0.59 ps | 0.60 ps | 0.48 ps | 0.49 ps |
| FWHM | 1.0 ps | 0.83 ps | 0.90 ps | 0.76 ps | 0.80 ps |

| **444 nm, 5K** | 0.30 mW | 1.0 mW | 3.0 mW | 10 mW | 18 mW |
|---|---|---|---|---|---|
| Exp decay | 0.54 ps | 0.59 ps | 0.56 ps | 0.53 ps | 0.51 ps |
| FWHM | 0.95 ps | 0.97 ps | 0.98 ps | 0.80 ps | 0.88 ps |

| **449 nm, 5K** | 0.10 mW | 0.30 mW | 1.0 mW | 3.0 mW | 10 mW | 18 mW |
|---|---|---|---|---|---|---|
| Exp decay | 0.88 ps | 0.83 ps | 0.85 ps | 0.85 ps | 0.77 ps | 0.70 ps |
| FWHM | 1.27 ps | 1.04 ps | 1.14 ps | 1.15 ps | 1.00 ps | 1.00 ps |

| 455 nm, 5K | 0.010 mW | 0.10 mW | 0.30 mW | 1.0 mW | 3.0 mW | 10 mW | 18 mW |
|---|---|---|---|---|---|---|---|
| Exp decay1 | | | | | 0.55 ps(L) | 0.58 ps | 0.55 ps |
| Exp decay2 | 11 ps (1) | 13 ps (1) | 20ps(2.5) | 12 ps(2) | 5.5 ps (4) | 3.9ps(9) | 3.3ps(11) |
| Exp decay3 | 78 ps (1) | 78 ps (1) | 78 ps (1) | 61ps (1) | 47 ps (1) | 29ps(1) | 25 ps (1) |

| 462 nm, 5K | 0.010mW | 0.10 mW | 0.30 mW | 1.0 mW | 3.0 mW | 10 mW | 18 mW |
|---|---|---|---|---|---|---|---|
| Exp grow1 | 1.0 | 0.73 | 0.97 | 0.74 | 0.76 | 1.0 (D) | 1.1 (D) |
| Exp grow2 | 8.8 (-1) | 4.2 (-1) | 4.9 (-1) | 3.3(-10) | | | |
| Exp decay1 | | | 31 (.7) | 7.1 (10) | 7.9 (10) | 6.4 (8) | 12 (5) |
| Exp decay2 | 157 (2) | 121 (1) | 119 (1) | 74 (1) | 44 (4) | 58 (1) | 67 (1) |
| Exp decay3 | | | | | 133(0.5) | | |

| 466 nm, 5K | 0.010mW | 0.10 mW | 0.30 mW | 1.0 mW | 3.0 mW | 10 mW | 18 mW |
|---|---|---|---|---|---|---|---|
| Exp grow1 | 1.0 | 1.4 | 1.0 /0.43 | 0.76 | 0.85 | 0.84(D) | 1.3 (D) |
| Exp grow2 | 13.4 (-1) | 11.6 (-1) | 6.8 (-4) | 3.3 (-7) | | | |
| Exp decay1 | | | 64 (2) | 26 (5) | 17 (4) | 6.9 (8) | 5.7 (6) |
| Exp decay2 | 202 (1) | 183 (1) | 162 (3) | 112 (3) | 92 (1) | 63 (1) | 60 (1) |

| 430 nm, 295 K | 0.30 mW | 1.0 mW | 3.0 mW | 10 mW | 18 mW | |
|---|---|---|---|---|---|---|
| FWHM | 0.87 ps | 0.92 ps | 0.92 ps | 0.87 ps | / | |
| EXP 1 | 0.44 ps | 0.48 ps | 0.46 ps | 0.53 ps | | |
| | | | | | | |
| 437 nm, 295 K | 0.10 mW | 0.30 mW | 1.0 mW | 3.0 mW | 10 mW | 18 mW |
| FWHM | 1.0 ps | 1.0 ps | 1.0 ps | 1.0 ps | 1.0 ps | 1.0 ps |
| Exp1 | 0.59 ps | 0.60 ps | 0.89 ps | 0.73 ps | 0.70 ps | 0.80 ps |

| 415 nm, 295 K | 0.30 mW | 1.0 mW | 3.0 mW | 10 mW | 18 mW |
|---|---|---|---|---|---|
| FWHM | / | / | 0.92 ps | 0.92 ps | 1.0 ps |
| Exp 1 | / | / | 0.48 ps | 0.38 ps | 0.52 ps |

| 422 nm, 295 k | 0.30 mW | 1.0 mW | 3.0 mW | 10 mW | 18 mW |
|---|---|---|---|---|---|
| FWHM | 0.70 ps | 0.87 ps | 0.92 ps | 0.87 ps | / |
| Exp 1 | 0.43 ps | 0.50 ps | 0.43 ps | 0.49 ps | / |

| 444 nm, 295 K | 0.10 mW | 0.30 mW | 1.0 mW | 3.0 mW | 10 mW | 18 mW |
|---|---|---|---|---|---|---|
| Exp 1 | 0.91 ps | 0.75 ps | 0.91 ps | 0.71 ps | 0.92 ps | 0.70 ps |
| Exp 2 | 59 ps | 38 ps | 27 ps | 8.54 ps | 6.40 ps | 4.60 ps |

| 464 nm, 295 K | 0.10 mW | 0.30 mW | 1.0 mW | 3.0 mW | 10 mW | 18 mW |
|---|---|---|---|---|---|---|
| Exp1 | / | / | 9.3 ps | 4.5 ps | 3.4 ps | 2.8 ps |
| Exp 2 | 20 ps | 22 ps | 21 ps | 18 ps | 15 ps | 15 ps |
| Exp3 | 86 ps | 86 ps | 64 ps | 62 ps | 63 ps | / |

## 4. Confusion over femtosecond HExEm decay and analysis by Forster-Dexter energy transfer model.

We have tried to use the Forster resonance energy transfer model[S1,S2] to analyze the observed femtosecond HExEm decay, yet confusions obtained are detailed below.

**Estimation of Forster resonance energy transfer (FRET) efficiency ($E$) and rate ($k_{ET}$).**

The measured quantum yield $Q_0$=~25% at 295 K for MeLPPP film, same as the reports in literature[S3,S4], from the PL spectrum integration ratio of 2:1 in Fig. 1(c), $Q_0$=~50% at 5 K.

Fitting the curve for $\lambda_{em}$=466nm at low $P$ in Fig. 2(f) gives $\tau_{rad}$= 200 ps for MeLPPP film at 5 K (see table above), then,

$\tau_0 = \tau_{rad} / Q_0$=200 ps /50%= 400 ps,

at $\lambda_{em}$=466 nm no ET takes place given it is at the bottom of DOS, thus the decay rate,

$$k_{rad\_466nm} = (k_f + \Sigma k_i) = \frac{1}{\tau_{rad}} = \frac{1}{200 ps} = 5.0 \times 10^9 \text{ sec}^{-1},$$

here $k_f$ is the radiative decay rate, $k_i$ are the rate constants of any other de-excitation pathway, including annihilation.

**Suppose** some short chain-segments existing in MeLPPP film with emission at $\lambda_{em}$=~442 nm at 5 K, thus we can reasonably assume its $\tau_{rad}$= 200 ps, and $\tau_0$= 400 ps. With the $\tau$= ~0.80 ps in Fig. 2(b), then the FRET efficiency,

$E$=1- $\tau/\tau_{rad}$=1- 0.80/200= **99.6%**,

**This value is too high to be acceptable! With this value a ten-step energy transfer can have a final efficiency, $(99.6\%)^{10}$= 96.0%. If this was true in an amorphous polymer film, thus the energy transfer efficiency is totally comparable to the photosynthetic system, where the proteins have much clearer stereo chemical configurational and conformational structures than the MeLPPP film used here, and in a common belief its energy transfer efficiency is just ~95% within several steps. Whereas up to date, nobody does really understand the real ET mechanism in the photosynthetic system[S5].**

Accordingly, the decay rate at $\lambda_{em}$=~442 nm should include an item of ET,

$$k_{rad-444nm} = k_{ET} + (k_f + \Sigma k_i) = \frac{1}{\tau} = \frac{1}{0.80 ps} = 1.25 \times 10^{12} \text{ sec}^{-1} \approx k_{ET} \gg 5.0 \times 10^9 \text{ sec}^{-1}$$

**Here the rate $k_{ET} = 1.25 \times 10^{12} \text{ sec}^{-1}$ is nearly the highest one in previous reports, unbelievable!**

**Estimation of Forster energy transfer distance (r):**

With Forster equation[S1,S2],

$$k_{ET} = (k_f + \Sigma k_i)(\frac{R_0}{r})^6 = \frac{1}{\tau_{rad}}(\frac{R_0}{r})^6$$

Here $R_0$ is the Forster radius, at which the ET efficiency is 50%.

With the data above,

$$k_{ET} = \frac{1}{200 ps}(\frac{R_0}{r})^6 = 1.25 \times 10^{12} \text{ sec}^{-1}$$

thus,

$$r = 0.398 R_0$$

For $\tau$ =0.40 ps – 1.0 ps for HExEm decay in Figs 2 & 3, the corresponding,

$$r = 0.355 R_0 \sim 0.414 R_0$$

**Estimation of $R_0$,**

The optical density (OD) is defined as, $OD = A = -\log\left(\frac{I}{I_0}\right) = \alpha d$

In Beer-Lambert law, $A' = -\ln\left(\frac{I}{I_0}\right) = \alpha' l$

So, $A' = A\ln(10) \approx 2.303 A$ and, $\alpha' = \alpha \ln(10) \approx 2.303 \alpha$

By $\alpha' = \frac{4\pi}{\lambda} k$, there is,

$2.303 \alpha = \alpha' = \frac{4\pi}{\lambda} k$, here $k$ is the extinction coefficient, and,

$$k = \frac{2.303}{4\pi}\alpha\lambda = \frac{2.303}{4\pi}\frac{A}{l}\lambda = 0.183 \frac{A}{l}\lambda,$$

By literature[S6,S7], for n=12 the MeLPPP oligomer /chain-segment has the 0-0 absorption peak at ~442 nm in solution at room temperature, which has similar shape as in Fig. 1(b), thus for estimation of Forster radius ($R_0$) at 5 K, set both absorption and PL 0-0 peaks at ~442 nm for the best spectral overlap, as shown in Fig. S4**a**.

1.) Normalize the PL spectrum as 1.00 by area, the result is shown in Fig. S4**a**.

2.) By $k = 0.183 \frac{A}{l}\lambda$ above, $l$=~120 nm (the typical thickness of MeLPPP thin film in our experiment), $A$ is the OD value measured (which is shown in Fig. 1 in the main text), work out the absolute extinction coefficient $k$, its max=0.80 at ~442 nm.

3.) The reported $\varepsilon$ value is $2.58\times10^5$ /M/cm at ~442 nm[S6,S7], which is used it to calibrate the $k$ spectrum obtained in 2 above, the resultant $\varepsilon$ curve in shown in Fig. S4**a**.

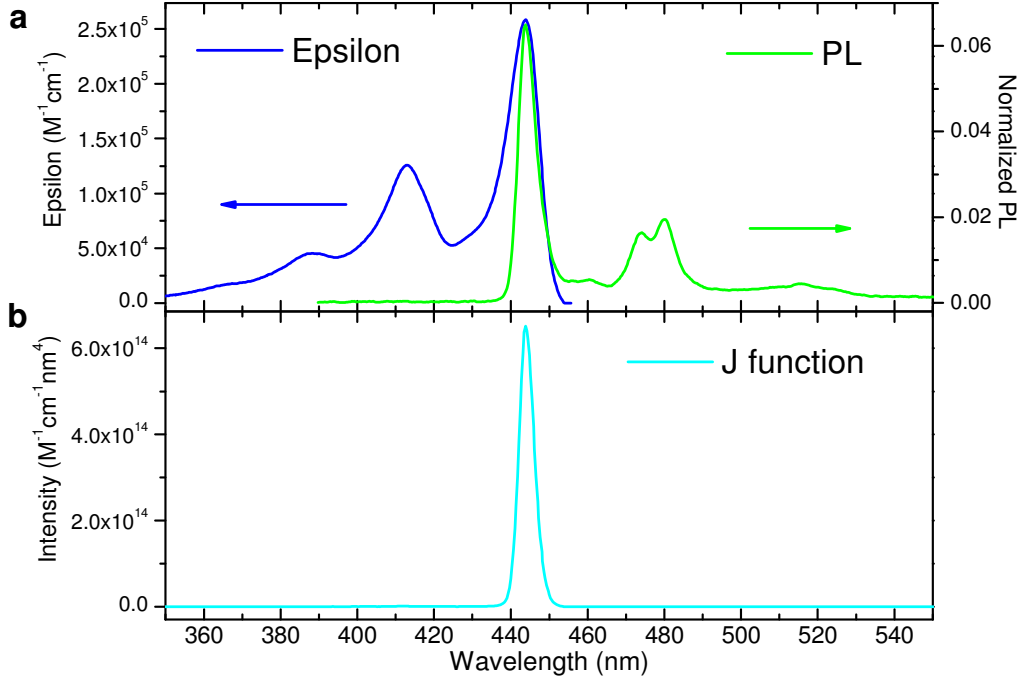

FIG. S4. Calculation of spectral overlap J-function for estimation of $R_0$. **a**. spectral overlap. **b**, calculated J-function.

4.) The calculated function of spectral overlap is shown in Fig. S4**b**. Doing the integration gives,

**J = 3.315×10$^{15}$ (M$^{-1}$cm$^{-1}$nm$^4$)**

With $R_0^6 = 8.79\times10^{-5}\left[k^2\eta_D n^{-4} J(\lambda)\right]$

and, $\eta_D$ = Q$_0$=50%, $k^2$=2, the orientation factor, n=1.90, the assumed refractive index,

we obtain $R_0^6 = 2.235\times10^{10}$, thus $R_0 =$ **5.31 nm**, this is a normal value for Forster radius.

then, $r = 0.398R_0$ **=2.11 nm**

And, $r = 0.355R_0 \sim 0.414R_0$ **=1.89 nm~2.20 nm**

Here the estimated Forster distance ($r$=~2.00 nm) is shorter than the exciton radius which extends to a whole chain by the conjugation effect. Assuming 140 pm length for C=C bound, then a repeat unit has about 6*140 pm= 840 pm= 0.84 nm, thus two repeat units have a length of 1.68 nm, this means, two excitons should be separated between two and three repeat units on a single chain. Also, this separation can not be an inter-chain distance given the side-group of a MeLPPP chain with ten C-C bonds prohibits this short distance (*r*=~2.00 nm). **Moreover, this short distance indicates that this is the realm of Dexter energy transfer which is more efficient than Forster below about 2.0 nm! However, we have not seen any previous report in a similar case using the Dexter model to deal with the excitonic singlet-singlet ET. If this was true, the photo-synthetic system can be understood by following this way.**

In our estimation $\kappa^2 =2$ is assumed (normally $0 \leq \kappa^2 \leq 4$). The ordinary $\kappa^2 =2/3$ is applicable when both segments are freely rotating and can be considered to be isotropically oriented during the excited state lifetime. If either chain-segment is fixed or not free to rotate, in our case, the fluorescent chain-segments in the pristine thin films of MeLPPP do not reorient on a femtosecond timescale that is faster than the energy transfer time observed in our experiment, then $\kappa^2=2/3$ will not be a valid assumption thus we use $\kappa^2 =2$ in our estimation. In most cases, however, even modest reorientation of the segments results in enough orientational averaging that $\kappa^2 = 2/3$ does not result in a large error in the estimated energy transfer distance due to the sixth power dependence of $R_0$ on $\kappa^2$. Even when $\kappa^2$ is quite different from 2/3 the error can be associated with a shift in $R_0$ and thus determinations of changes in relative distance for a particular system are still valid, for example, with $\kappa^2 = 2/3$, we obtain $R_0^6 = 0.745 \times 10^{10}$, thus $R_0 =$ **4.42 nm**, and $r = 0.355 R_0 \sim 0.414 R_0$ =**1.57 nm~1.83 nm< 2.0 nm.**

**Considerations by Dexter energy transfer**

The Dexter energy transfer is another fundamental phenomenon in photophysics. The difference between Forster and Dexter mechanism include (1) Dexter mechanism involves the overlap of wavefunctions so that electrons can occupy the other's molecular orbitals. (2) The reaction rate constant of Dexter energy transfer sharply decreases while the distance between Donor and Acceptor increase and the distance is generally smaller than 1.0 nm. (3) The Dexter mechanism can usually be applied to produce the triplet state of some molecules of interest. The special case of exchange-triplet-triplet annihilation-can "push" the electron to upper singlet states by exchanging the electrons of two triplet molecules.

The rate constant of exchange energy transfer is given by,

$$k_{Dexter} = kJ \exp\left(\frac{-2R_{DA}}{l}\right)$$

Here *J* is the normalized spectral overlap integral, same as above. The term "normalized" means making the absorption spectra and the emission spectra on the same scale and has the same highest level, **k** is an experimental factor, **R_{DA}** is the distance between Donor and Acceptor chromophores, and *l* is the sum of van der Waals radius. The rate constant of exchange energy transfer decays steeply because of its intrinsic exponential relationship, which is the reason that the exchange energy transfer is also called the short-range energy transfer and the Forster mechanism is called the long-range energy transfer.

In more details, the Dexter energy transfer is a process that the donor and the acceptor exchange their electron. In other words, the exchanged electrons should occupy the orbital of the other party. Hence, besides the overlap of emission spectra of Donor and absorption spectra of Acceptor, the exchange energy transfer needs the overlap of wavefunctions. In the popular words, it needs the overlap of the electron cloud. The overlap of wavefunctions also implies that the excited donor and ground-state acceptor should be close enough so the exchange could happen. However, this can not be applied to the case here for MeLPPP, the overlap of orbital and wavefunction straightforwardly means the effective dislocation of π-conjugation on a chain, this forms a chromophore for an exciton to dwell, and not allowed for two excitons on one chromophore. Therefore, the Dexter ET model can not be used here for the femtosecond HExEm decays.